\def \SAIT #1 #2 {{\em Mem.\ Soc.\ Astron.\ It.\/} {\bf #1}, #2}
\def \MESS #1 #2 {{\em The Messenger\/} {\bf #1}, #2}
\def \ASTRNACH #1 #2 {{\em Astron. Nach.\/} {\bf #1}, #2}
\def \AAP #1 #2 {{\em Astron. Astrophys.\/} {\bf #1}, #2}
\def \AAL #1 #2 {{\em Astron. Astrophys. Lett.\/} {\bf #1}, L#2}
\def \AAR #1 #2 {{\em Astron. Astrophys. Rev.\/} {\bf #1}, #2}
\def \AAS #1 #2 {{\em Astron. Astrophys. Suppl. Ser.\/} {\bf #1}, #2}
\def \AJ #1 #2 {{\em Astron. J.\/} {\bf #1}, #2}
\def \ANNREV #1 #2 {{\em Ann. Rev. Astron. Astrophys.\/} {\bf #1}, #2}
\def \APJ #1 #2 {{\em Astrophys. J.\/} {\bf #1}, #2}
\def \APJL #1 #2 {{\em Astrophys. J. Lett.\/} {\bf #1}, L#2}
\def \APJS #1 #2 {{\em Astrophys. J. Suppl.\/} {\bf #1}, #2}
\def \APSS #1 #2 {{\em Astrophys. Space Sci.\/} {\bf #1}, #2}
\def \ASR #1 #2 {{\em Adv. Space Res.\/} {\bf #1}, #2}
\def \BAIC #1 #2 {{\em Bull. Astron. Inst. Czechosl.\/} {\bf #1}, #2}
\def \JSQRT #1 #2 {{\em J. Quant. Spectrosc. Radiat. Transfer\/} {\bf #1}, #2}
\def \MN #1 #2 {{\em Mon. Not. R. Astr. Soc.\/} {\bf #1}, #2}
\def \MEM #1 #2 {{\em Mem. R. Astr. Soc.\/} {\bf #1}, #2}
\def \PLR #1 #2 {{\em Phys. Lett. Rev.\/} {\bf #1}, #2}
\def \PASJ #1 #2 {{\em Publ. Astron. Soc. Japan\/} {\bf #1}, #2}
\def \PASP #1 #2 {{\em Publ. Astr. Soc. Pacific\/} {\bf #1}, #2}
\def \NAT #1 #2 {{\em Nature\/} {\bf #1}, #2}

\documentstyle{memsait}
\input epsf.sty
\begin{opening}
\title{THE PROPAGATION OF LIGHT POLLUTION IN DIFFUSELY URBANISED AREAS} 
\author{PIERANTONIO CINZANO}
\institute{Dipartimento di Astronomia, Universit\`a di Padova,
vicolo dell'Osservatorio 5,\\  I-35122 Padova, Italy\\
email: cinzano@pd.astro.it}
\date{} 
\end{opening}

\begin{document}

\oddpagefooter{}{}{} 
\evenpagefooter{}{}{} 
\ 
\bigskip

\begin{abstract}
The knowledge of the contribution $b_d(d)$ to the  artificial
sky luminance  in a given point of the sky of a
site produced by the sources beyond a given distance $d$ from it
is important to understand the behaviour of light pollution in diffusely urbanized areas and to estimate which fraction
of the artificial luminance would be regulated by norms or laws limiting the light wasted upward within protection areas of given radii. 

I  studied the behaviour of  $b_d(d)$ 
constructing a model for the propagation of the light
pollution based on the modelling technique introduced by
Garstang which allows to
calculate the contribution to the artificial luminance in a
given point of the sky of a site of given altitude above sea level,
produced by a  source of given emission and geographic position.
I obtained $b_d(d)$ integrating
the contribution to the artificial luminance from
every source situated at a distance greater than $d$.
I also presented an analitical expression for $b_d(d)$ depending mainly from one parameter, a core radius, well reproducing model's results. The  artificial
sky luminance $b_d(d)$ produced in a given point of the sky of a
site from the sources situated at a distance
greater than $d$ from the site decrease in diffusely urbanized areas as $ d^{-0.5}$ i.e. much slowler than the artificial luminance of a single source which decrease as $d^{-2.5}$.
 In this paper I present the results for $b_d(d)$ at some Italian Astronomical Observatories. In a diffusely urbanised territory
the artificial sky luminance   produced by sources located at large distances from the site is not negligible due at the additive character of light pollution and its propagation at large distances.
Only when the core radius is small, e.g. for sites in the inner outskirts of a city, the sky luminance  from sources beyond few kilometers is negligible. The radii of protection zones around Observatories needs to be large in order that prescriptions limiting upward light be really effective.

\end{abstract}

\section{Introduction}
Light pollution is a quantity characterized by additivity and propagation at large distances. 
Walker (1970, 1973) showed that the light coming from a big city can pollute the sky at a great distance from it. Garstang (1989b) calculated with numerical models the artificial sky luminance in some quite isolated Astronomical Observatories of worldwide interest showing the frequent presence of non negligible pollution coming from light propagating there from very far sources. Already Bertiau et al (1973) showed the dramatic impact of additive behaviour of artificial light scattered by atmospheric particles and molecules in determining the sky luminance and the related limiting magnitude loss.

The artificial sky luminance produced in a site from a
single source, like a city, decreases rather rapidly with the distance from the source. 
In fact roughly $b\propto d^{-2.5}$ (Walker 1973; see also Garstang 1991a). This slope might wrongly suggest 
 that in 
territories where there aren't big cities able to produce light pollution to great distances only the  sources of light pollution situated in the
neighbourhood of a site are responsible of  the artificial  luminance of its sky. This is not true in diffusely populated areas where there aren't big cities but a spread-out myriad of small cities because the artificial sky luminance
produced from each source sums together. In this case the effects depend strictly on the sources distribution.
When a territory is so diffusely populated that the distribution of sources can be considered a continuum, circular sections with constant thickness and increasing radius centered on the site contain an increasing number of sources so that the diminution of their effects on the sky at the site due to the increased average distance is somehow counterbalanced by the growth of the  number of sources. So the global effects of all sources beyond a given distance decrease with this distance much slower than the effect of a single source.
This is a fundamental phenomenon in areas, like as
an example the Padana plain in Veneto (Italy), where the population is
distributed in a myriad of cities and towns in such way to cover almost all the entire territory. 
This means that  zones of territory even very far 
from the site can contribute remarkably to the
luminance of its sky. This implies that laws
for the protection of the sky of a site must be able to act
on polluting sources at remarkable
distances from the site, otherwise their efficacy would be scarce. 
			
In this paper I demonstrate, both with simple analytical and detailed numerical modeling for light pollution propagation, that contributions to the zenith sky luminance from areas beyond a given distance decrease slowly in diffusely urbanized areas with increasing distance. I also present an analytic formula working for diffusely populated territories and  results for some Italian Observatory sites. 
In section 2 the analytic formula for the sky luminance produced by sources beyond a circular area of given radius is presented and discussed. In section 3 the detailed modelling technique is illustrated. In section \ref{comp} the results are compared with those obtained with other laws. In section 5 results for some Italian Observatories are presented. The section 6 contains my conclusions.

\section{Elementary theoretics}
In order to understand the behaviour of light pollution in diffusely urbanized areas, let's
calculate the zenith sky luminance at a
site produced from the sources
beyond a given distance $d$ from it. 
Let's
assume that in a territory the population is distributed
homogeneously with a density of $p$ inhabitants for unit of
surface. This happens, as an example, in the Veneto's plain (Italy) where the population is distributed in a myriad of cities and
towns so it covers almost all the
territory. Let's also assume that a law of the type $b=p~f(r)$ is valid
there which gives the artificial luminance $b$ at the zenith produced by a source city of given population $p$ in function of the
distance $r$ from it and
that this law can be applied to each area of
territory of population $p$.

The area of an infinitesimal circular section of
thickness $dr$ at the distance $r$ from the site and centered on it, is
$2\pi r dr$ and its total population is $2\pi p_0 r dr$, where $p_0$ is the population for surface units in the considered territory. 
The
sky luminance produced at the zenith of the site from this infinitesimal circular section 
is 
 $2\pi p_0 r f(r)dr$.
In order to compute the sky luminance $b_d$ produced by all the sources outside the distance $d$ we need to integrate the last expression between $d$ and infinity : 
\begin{equation} 
\label{int}
b_d =2\pi p_{0} \int^{\infty}_{d} f(r) r dr  
\end{equation} 
Even without performing the computation of the integral, readers can see that on increasing the distance from the site $b_d$ decreases slower than the function  $f(d)$. 

In order to compute the integral (\ref{int}) we need a Law giving the sky luminance at the zenith.  In order to became simple,
we will assume $f(r) \propto r^{-\nu}$ which is an extension of Walker Law. The exponent $\nu$ depends on the aerosol content of the atmosphere through its optical thickness and on the zenith and azimuth angles of the direction of observation (Joseph et al. 1991; Garstang 1986). We assumed at zenith $\nu \approx 2.5$ which is valid in the range between 1 and 30 km for an optical thickness of $\sim 0.25$ (Joseph et al. 1991, fig. 5 - 6) giving a vertical extinction of $k_V\sim0.3$ mag V. Garstang (1986, 1989) showed that a power law relation between the artificial sky luminance and the distance $r$ of its source is not exact, the exponent $\nu$ of Walker Law becoming larger for increasing distance $r$. The effects of Earth curvature contribute to this tendency. Nevertheless, $\nu=const.$ {\it everywhere} is an adequate approximation for the purposes of this computation. Other propagation laws, like e.g. 
the Treanor Law (Treanor 1973), can be used in (\ref{int}), as well more sophisticated computations like that I used in section \ref{results}, which will be described in section \ref{model}.

With $f(r) \propto r^{-2.5}$ the
integral solved gives (Cinzano 1997):
\begin{equation}
\label{eq7}
b_{d}\propto p_{0} \times d^{-0.5}
\end{equation} 
This law expresses the contribution $b_d$ to the  total artificial luminance at the zenith in
a site produced from all the territory situated outside a
given distance $d$. 

It can be improved 
taking in  account two phenomena: (i) when the distance from a
site is under a given value that we
can call " Core radius " the uniformity of the distribution of the
sources stops because the scale comes down to the level
of the irregularities in the distribution itself. Therefore $b_d$ deviates from the
expression (\ref{eq7}) and tends to become constant in the
neighbourhood of the site where there aren't light sources. (ii) the
curvature of the Earth diminishes in a non negligible way the contribution of the areas
beyond about 80 km. 
In order to mimic approximately the behaviour produced by (i), I corrected the
expression (\ref{eq7}) inserting a "core". In order to mimic approximately the behaviour given by (ii), I inserted in the
expression (\ref{eq7}) a little correction factor $k(d)$ taking in account the diminution of contribution from distant sources. If the law is not applied beyond about 100  km from the
site, in first approximation k can be considered a constant which subtracts the overestimated contribution from sources outside this distance. Taking into account both phenomena we can write:
\begin{equation}
\label{eqlaw}
b_d\propto p_{0} \left( \left( d_{c}^{\alpha}+d^{\alpha} \right)^{-0.5/\alpha}
-k \right)
\end{equation} 
where $d_c$ is the core radius, $\alpha$ is a parameter giving the
shape of the curve in the core, where $d\leq d_{c}$.
For $d>>d_{c}$ the slope of $b_d$ is like expression (\ref{eq7}) and for
$d<< d_{c}$ it becomes
constant. The core radius
 $d_{c}$ isn't the distance from
the site of the nearest source but it is the scale distance under which the uniformity of population distribution vanishes. 

Normalizing the formula (\ref{eqlaw}) so that $b=1$ for $d=0$, we obtain:
\begin{equation} 
\label{bcontrib} 
b_d = \frac{ \left( \left( d_{c}^{\alpha}+d^{\alpha} \right)^{-0.5/\alpha}
-k \right)}{d^{-0.5}_{c}-k} \approx  \left(1+\left( 
\frac{d}{d_{c}} \right)^{\alpha} \right)^{-0.5/\alpha}  - kd_{c}^{0.5}
\end{equation}
because $k<<d_{c}^{-0.5}$.  Some applications of this formula will be presented in section \ref{comp}. The core radius is strictly depending on the distribution of sources around the site. The slope parameters and the $k$ coefficient are less sensitive and typical values are respectively $\alpha=3$ and $k=(120)^{-0.5}$.

\section{Detailed modelling} 
\label{model}
In order to study with more detail  the  artificial
sky luminance $b_d(d)$ produced in a given point of the sky of a
site from the sources situated at a distance
greater than $d$ from the site, I 
applied a model for the propagation of the light
pollution based on the modelling technique introduced by
Garstang (1986, 1987, 1988, 1989a, 1989b, 1989c, 1991a, 1991b, 1991c, 1992, 1993, 1999). The model allows to
calculate the contribution to the artificial luminance in a
given position on the sky of a site of given altitude above sea level,
produced by a  source of given emission and geographic position.
I obtained $b_d(d)$ integrating
the contribution to the artificial luminance of
every source situated at a distance greater than $d$. Readers are referred to the papers cited above for a detailed description and discussion of the models. Here I will  describe their outline.
			
For
every infinitesimal volume of atmosphere along the line of sight, I 
calculated the  illuminance  produced directly by the
source and that produced by the light scattered there from
molecules and aerosols, estimated
according to the method of Treanor (1973) extended by Garstang (1986, 1989). Then I computed the quantity of light scattered in direction of the observer by the
molecules and the aerosols in the infinitesimal volume.
Integrating along the line-of-sight, I obtained the
artificial luminance of the sky. In photometrical bands different from the visual band, as the astronomical B band, the photon radiance can be obtained. If required, they can be transformed in brightnesses in $mag/arcsec^{2}$ with Garstang (1989) formulae.

As Garstang(1986, 1989a), I assumed  that the distribution of the
upward emission of light from a city is axisymmetric, i.e. it depends only on the
angle $\theta$ with the vertical, and that it is expressed by a function $f(\theta)$ that will be discussed in section \ref{comp}.
I assumed that the lighting habits are similar in
all the cities of the considered territory, and  than the quantity
of light wasted from the night-time external lighting system
of a city is proportional to its population (e.g. Walker 1988). Falchi and Cinzano(1999) analyzing some night-time satellite images of Italy pointed out that the  upward emission of italian cities might depend on a power of 0.8 of their population. Nevertheless in the range of population considered here the direct proportionality between the upflux and the population is a good approximation for the purposes of this paper.
I assumed that differences from the mean were casually distributed in the territory. Considering economically
homogenous zones, not larger than 120-150 km, I had not to insert in the
calculation a coefficient of city development connected with the geographic position as Bertiau et al. (1973). The population data refers to the year 1995 and geographical positions and radii of cities refers to 1991. They were provided by the Istituto Nazionale di Statistica (ISTAT).
I have considered as point sources
the cities when the line of sight did not approach them closer than 12
times their radius and I have used in the other cases a  seven points approximation (Abramowitz and Stegun
1964).
I assumed that the density of molecules
and aerosols decreases exponentially with the
altitude. The angular scattering function of the
atmospheric aerosols was given by
Garstang (1991b) interpolating measurements from
McClatchey et al. (1978).  

The aim of the model is the computation of the artificial luminance in a site not far from
very populated areas and 
not the computation of  the small luminance produced 
by far sources in very good international observatory sites like e.g. Garstang (1989b), so I have neglected to take in account
the Earth curvature. The inclusion of earth curvature causes luminance contributions from distant cities to fall off more rapidly with distance than for a flat Earth model (Garstang 1989a). The effects of the land curving on the
luminance near the zenith are of the order of
2 percent for sources outside 50 km  and can reach 35-40 percent at about 100km (Garstang 1989a).
In the applications that have
been made, a predominant fraction of luminance was
produced within the first 60 km from the
site and less than 5 percent outside about 90 km, so the choice to neglect curvature is adequate.  

In the calculation I have not taken in account that mountains
can shield part of the light of a source to the 
atmospheric particles situated along the line of sight of the
observer at the site. 
Mountains  between source and observatory  shield the light emitted from the source with an angle less than  $\theta =
arctg\frac{H}{p}$ where  H is the height of the mountain and p its
distance from the source. The ratio between the
luminance in the shielded and not shielded cases is given, in first approximation, by the ratio between the number of
particles  illuminated in the two cases: 
\begin{equation}  \frac{b_s}{b_{ns}}\approx\frac{  
\left[ \int^{\infty}_{Hq/p} N_a(h) dh \right] \sigma_a(\psi)+  
\left[ \int^{\infty}_{Hq/p} N_m(h) dh \right] \sigma_m(\psi)  
}  
{  
\left[ \int^{\infty}_{0} N_a(h) dh \right] \sigma_a(\psi)+  
\left[ \int^{\infty}_{0} N_m(h) dh \right] \sigma_m(\psi)  
}  
\end{equation}
where q is the distance of the site from the source, $N_a(h)$, $N_m(h)$, $\sigma_a(\psi)$ and $\sigma_m(\psi)$ are respectively the number of particles of aerosol and molecules at the altitude h and their angular scattering sections and roughly $\psi \approx \pi-\theta$.
This expression show that, taken in account the vertical size of the atmosphere in respect
the height of mountains, shielding has a not negligible
effect only when the source is  quite near to the 
mountain and both are very far from the
site: $ \frac{q}{p} >> \frac{a}{H}$. This  condition in the territories considered could apply for few sources only.
I neglected the effects of the 
the Ozone layer  and the volcanic powder studied by Garstang (1991). This kind of model has been already widely tested by Garstang (see ref. above). I successfully tested my set of models in Italy comparing the predicted sky luminance distribution in some italian sites with measurements. Results of tests were reported elsewhere (Cinzano 1999). In next section I will also compare  the $b_d$ obtained with the model for a test site and that obtained with the Treanor Law.

\section{Comparison tests.}
\label{comp}
I
compared  the expression (\ref{bcontrib})  with the normalized $b_d(d)$ curve obtained for the Astronomical Observatory of Ekar  with the  models described above (see sec. \ref{results}).
The surrounding
territory of the Veneto plain is in fact studded with
many cities and small towns in nearly continuous
way. 
\begin{figure}
\epsfysize=8cm 
\hspace{3.5cm}\epsfbox{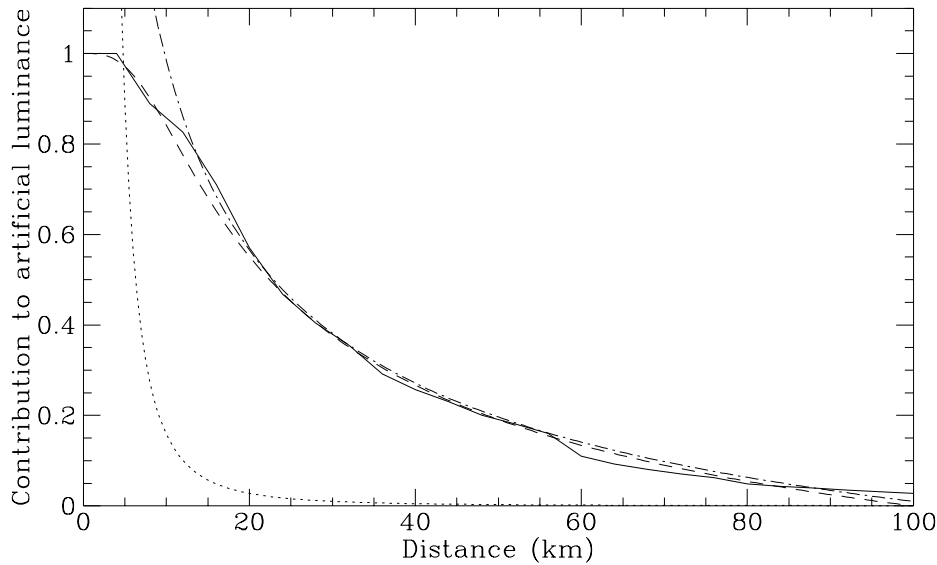} 
\caption{Comparison between $b_d(d)$ curves. The solid curve is the model prediction and the dashed curve is the best fit of expression (\ref{bcontrib}). The Walker Law ($b\propto d^{-2.5}$) (dotted curve) and the $b\propto d^{-0.5}$ Law (dot-dashed curve) were also plotted.}
\label{comp1}
\end{figure}
The results are presented in figure \ref{comp1}. The solid curve is the model prediction and the dashed curve is the best fit of expression (\ref{bcontrib}). The Walker Law ($b\propto d^{-2.5}$) (dotted curve) and the $b\propto d^{-0.5}$ Law (dot-dashed curve) were also plotted.
I  obtained a good
agreement for $\alpha=3$, $d_{c}=10$ km and $k$=0.06. This value of
the Core radius corresponds roughly to the distance of the
Observatory from the surrounding plain. 
The choice of the
parameter $\alpha$ does not influence strongly 
the results because its
effect is non negligible only  for $d\leq d_{c}$ and the
contribution of the zones inside this distance is rather small compared with contributions coming from greater distances.

I also compared  the normalized $b_d(d)$ curve predicted for Mount Ekar Astronomical Observatory with the  curve
obtained 
calculating the contribution one by one of all the sources of
the surrounding territory with the  Treanor law (1973):
\begin{equation}  b= b_{0} I_{0} \left(
\frac{A}{X}+\frac{B}{X^{2}} \right)~~e^{-kX}  \end{equation} 
where A,B,k, $I_0$, $b_0$ are constant and X is the distance of the site from the source.
 The Treanor law  refers to a simple model with homogenous atmosphere, the heights of 
scattering particles are small respect the distance between the 
site and the source, and the scattering is assumed to be limited to a
cone of small angle around the direction of the incident light. These hypotheses are justified in first
approximation from the limited scale height of
atmospheric particles and from the characteristics of
forward scattering of the aerosols. The model
neglects the scattering of higher order than the second. The Treanor law, even if based on a simple model,
fitted well the measurements  carried
out by Bertiau et al (1973) in Italy. The term $1/X^{2}$  give the contribution to the 
illuminance of the  atmospheric particles along the vertical
column at the site produced by light coming directly
from the source, the term $1/X$  gives the
contribution of the  light scattered once, the term $e^{-kX}$
takes into account the extinction of the light along its way. In applying this law I  considered still valid
the calibration of the ratio B/A and the
coefficient k from Bertiau et al (1973) in B band because these
depend only on the mean conditions of the atmosphere
in clear nights which I supposed unchanged from
1973, neglecting the seasonal variations and the
effects of the changes in atmospheric pollution. 
\begin{figure}
\epsfysize=8cm 
\hspace{3.5cm}\epsfbox{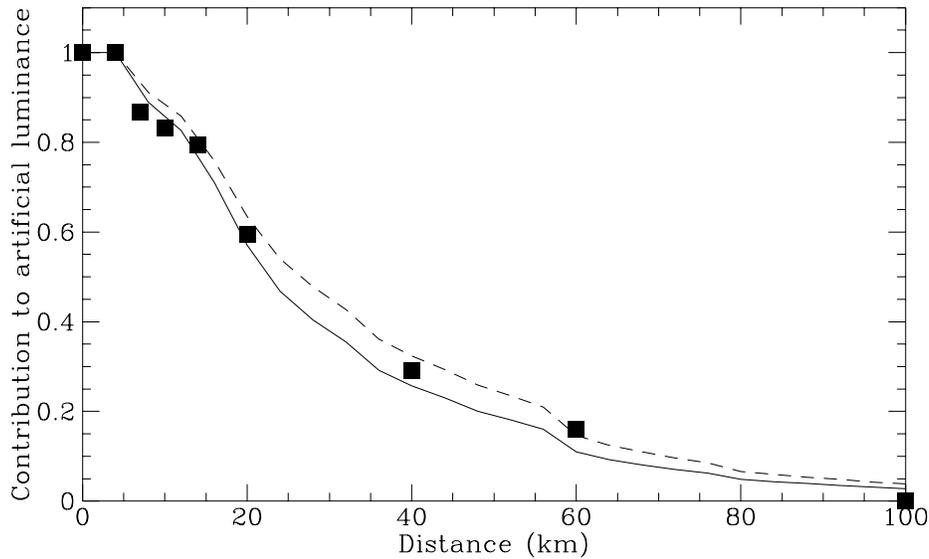} 
\caption{ Prediction of $b_d(d)$ obtained for Mount Ekar Astronomical Observatory in B band (solid curve) and in V band(dashed curve) with the  model  and in B band with the  Treanor Law (squares).}
\label{fig2}
\end{figure}
The effect of the altitude H of
the observatory is to cut all the light that in a site on the sea
level would come from particles situated along the
vertical at a lower height than the altitude H of the site. In a rough approximation, the luminance is overestimated for a fraction given by the ratio between the number of scattering 
in the
vertical column over the site in the case of altitude H
and in the case of altitude zero. 
\begin{equation}  \frac{b_H}{b_{0}}\approx\frac{n_{h}}{n_{0}}=\frac{  
\left[ \int^{\infty}_{H} N_a(h) dh \right] \sigma_a(90^{\circ})+  
\left[ \int^{\infty}_{H} N_m(h) dh \right] \sigma_m(90^{\circ})  
}  
{  
\left[ \int^{\infty}_{0} N_a(h) dh \right] \sigma_a(90^{\circ})+  
\left[ \int^{\infty}_{0} N_m(h) dh \right] \sigma_m(90^{\circ})  
}  
\end{equation}
where  $N_a(h)$, $N_m(h)$, $\sigma_a(\psi)$ and $\sigma_m(\psi)$ are respectively the number of particles of aerosol and molecules at the altitude h and their angular scattering cross sections.
With the expressions and the parameters in Garstang (1986)
I estimate that the effect of the altitude diminishes
the values of approximately 20\% for  the
observatory. However, all the
contributions are affected in the same
way from the minor number of atmospheric particles, so that, in first approximation,  relative contributions do not need any correction.

Figure \ref{fig2} shows the results of the computations of $b_d(d)$ with Treanor Law (squares) and the model predictions in B band (solid curve) and in V band (dashed curve).
Differences between the curves in V and B bands are likely to be of the same order as differences produced by fluctuations in atmospheric conditions. Measurements of Walker (1977, table II, col (2))in V and B bands at Salinas are almost indistinguishable from 10 km to 30km. The Treanor Law (valid for B band) and the Walker Law (obtained for V band) were found in fairly good agreement (Walker 1977). The differences on the propagation of light pollution in B band and in V band are shown in Cinzano \& Stagni (1999). 
\begin{figure}
\epsfysize=6cm 
\hspace{1.2cm}\epsfbox{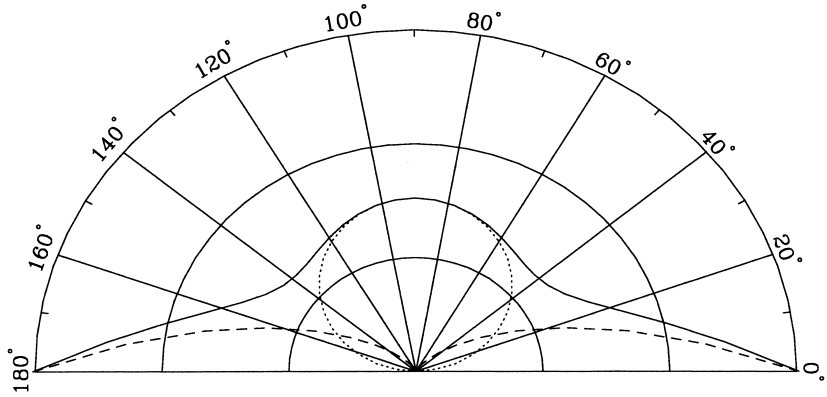} 
\caption{Mean upward emission function of cities from Garstang (1986) with G=0.15, F=0.15.}
\label{fig3d}
\end{figure}

\begin{figure}
\epsfysize=6cm 
\hspace{1.5cm}\epsfbox{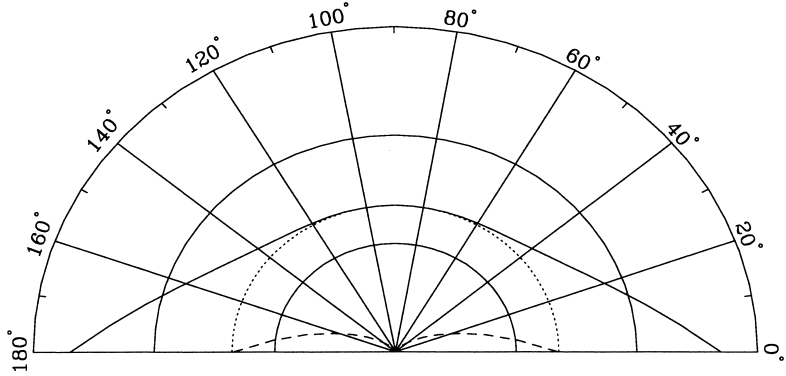} 
\caption{Mean upward emission function of cities: alternative form in eq. \ref{fnew}.}
\label{fig3e}
\end{figure}

\begin{figure}
\epsfysize=8cm 
\hspace{3.5cm}\epsfbox{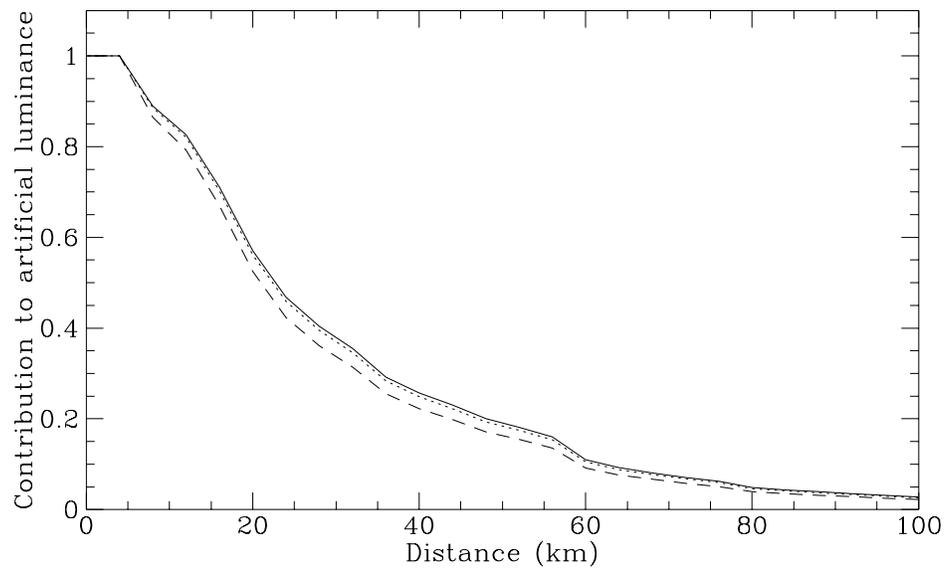} 
\caption{$b_d(d)$ curve obtained with the function (\ref{fclassic}) (solid curve),   the function (\ref{fnew}) (dotted curve),  and a constant intensity (dashed curve).}
\label{fig3g}
\end{figure}

The average upward emission function of cities $I(\theta)$ was never studied  so far in Italy or elsewhere. Garstang in all his papers cited here used successfully a semi-empirical function:
\begin{equation}
\label{fclassic}
I(\theta)=q\frac{1000}{2\pi} \left[ 2G(1-F) \cos \left( \frac{\pi}{2}-\theta \right) + 0.554 F \left( \frac{\pi}{2} -\theta \right)^{4} \right]
\end{equation}
where $q$ is a calibration constant unimportant in the normalized computation of this paper and $I(\theta)$ is in lm/sr.
This function was implicitly tested by Garstang with many comparison between model predictions and measurements.
I compared the $b_d(d)$ curve obtained with three different average upward emission functions $I(\theta)$. The first, shown in figure \ref{fig3d}, was the previous eq. \ref{fclassic}, with G=0.15 and F=0.15.
Note that G and F must be considered here only shape parameters without the meaning given by Garstang (1986) of, respectively, direct and reflected light ratios because  emission at low angles above the horizon can come both from direct emission by luminaires and from reflection by lighted surfaces (horizontal and vertical) and emission near the zenith can come both from reflexion by surfaces and from direct emission by unshielded luminaires.
The second function, shown in figure \ref{fig3e}, was the sum of a constant intensity and a power law:
\begin{equation}
\label{fnew}
I(\theta)=q\frac{1000}{2\pi} \left[ G(1-F) + 0.554 F \left( \frac{\pi}{2} -\theta \right)^{4} \right]
\end{equation}
with G=0.3 and F=0.075.
This function assumes an higher emission at intermediate angles in respect to the function \ref{fclassic}. As third function I used a constant intensity $I(\theta)=const$.
In figure \ref{fig3g} I plotted the $b_d(d)$ curve obtained with the function (\ref{fclassic}) (solid curve),  the function (\ref{fnew}) (dotted curve), and the constant intensity (dashed curve). Differences in $b_d(d)$ shape are little even for the case of constant intensity.  

\begin{figure}
\epsfysize=8cm 
\hspace{3.5cm}\epsfbox{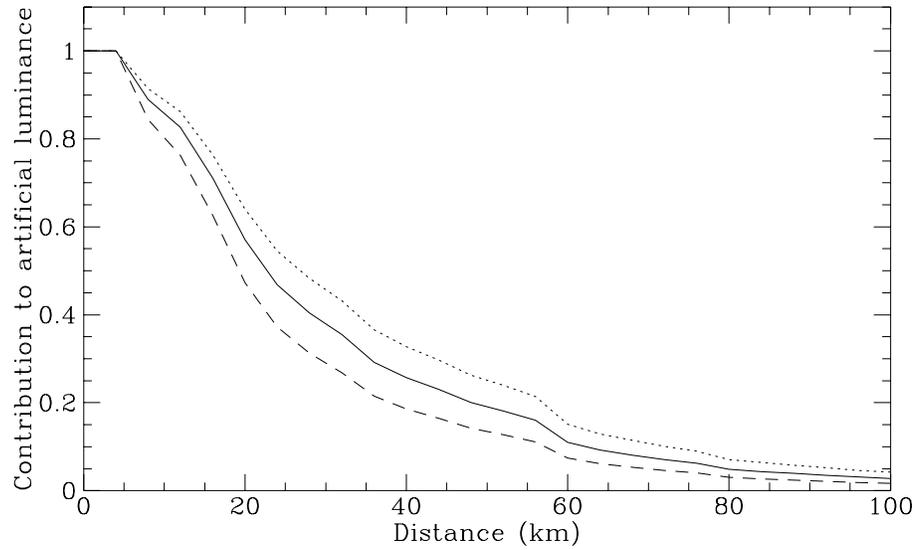} 
\caption{$b_d(d)$ curve obtained for Mount Ekar Observatory for K=1 (solid curve), K=0.5 (dotted curve) and K=2 (dashed curve).}
\label{fig3c}
\end{figure}
The propagation of light pollution depends on atmospheric clarity, i.e. on the aerosol content of the atmosphere, which in the models is expressed by a clarity parameter K (Garstang 1986, 1987, 1988, 1989a, 1989b, 1989c, 1991a, 1991b, 1991c, 1992, 1993, 1999). In computing $b_d(d)$ curve I assumed {\it typical clean air at sea level}, i.e. K=1 as defined by Garstang (1986). A clarity K=0.5 would correspond to very clean air.
So results in this paper refers to clean atmosphere not to mean atmospheric conditions at the studied sites, which are difficult to define. In figure \ref{fig3c} I compared the $b_d(d)$ curve obtained for Mount Ekar Observatory for K=1 (solid curve), K=0.5 (dotted curve) and K=2 (dashed curve).  An increase in aerosol content produces a steeper decrease of $b_d(d)$ for increasing d. This is due at the increase of scattering which produces a stronger extinction. This was already shown by Garstang (1986) for the city of Denver, where on increasing K the artificial luminance contribution decreases outside the city while increases inside the city.
A comparison between curves of figure \ref{fig3c} and figure \ref{fig2} shows that the model for K=2 does not agree with the curve obtained with the Treanor Law as calibrated with measurement in Italy (Bertiau et al. 1973). So atmospheric conditions at which Treanor Law was calibrated seems to correspond to an aerosol content $K\sim1$.
In table 2 I will show for each Observatory the vertical extinction in V and B band predicted by  the model (Garstang 1991) for K=1 and K=2. The horizontal daylight visibility, defined as the distance at which black object would show a luminance (due to scattered light between the observer and the object) of 0.98 of the luminance of horizon behind the object, computed as Garstang (1991), is $\Delta x=48$ km  for $K=0.5$,  $\Delta x=26$ km  for $K=1$ and $\Delta x=14$ km for $K=2$ at sea level.

\subsection{Results}
\label{results}
I computed $b_d(d)$ with the  detailed modelling described in section \ref{model} for the  Italian Observatories listed in table 1.   
Table 1  shows for each Observatory the adopted geographical position and altitude above the sea level. 

\vspace{1cm} 

\centerline{\bf Tab. 1 - Adopted geographic positions of Observatories}
\begin{table}[h]
\hspace{0.8cm} 
			\begin{tabular}{|l|r|r|r|}
			\hline
		Observatory  & Longitude & Latitude & Height\\
		&   $^{\circ}$ ~~$'$ ~~$''$  & $^{\circ}$ ~~$'$ ~~$''$  & m.s.l.\\
			\hline
Mount Ekar Obs. (Asiago)& 11 34 18  & 45 50 36  & 1350\\
Bologna Univ. Obs. (Loiano)& 11 20 ~0  & 44 15 23 & 714\\
Brera-Milan Ast. Obs. (Merate)& 9 25 42  & 45 41 58  & 330\\
Catania Obs. Stellar Sta. (Serra La Nave)& 14 58 24   &   37 41 30   & 1735\\   
Collurania Ast. Obs. (Teramo)&13 44 ~0  & 42 39 30  & 388\\
Chaonis Obs. (Chions)& 12 42 42  & 45 50 36  & 15\\
``G. Ruggieri'' Obs. (Padova)& 11 53 20    &  45 25 10   & 20\\
S. Benedetto Po Ast. Obs. (S.Benedetto Po)&10 55 10 &45 ~3 ~4 & 1.\\
\hline
\end{tabular}
\end{table}

 As already discussed, results were obtained for conditions of  clean atmosphere (i.e. assuming for the Garstang (1989) clarity parameter $K=1$). Table 2  shows for each Observatory the  vertical extinction in V and B band predicted by  the model (Garstang 1991)for K=1 and K=2. 

The $b_d(d)$ curves in figures are normalized to the total artificial sky luminance at the zenith of the site. They are computed for light in the photometric band of eye sensitivity and they can be considered also valid for light in the astronomical V band. The curves are computed every 4 km and linearly interpolated in the plots. A better space resolution would be sometime confusing because the city position data do not give the exact position of light baricenter but only the position of a geographic reference point in the center of the city.
\begin{figure}
\epsfysize=8cm 
\hspace{3.5cm}\epsfbox{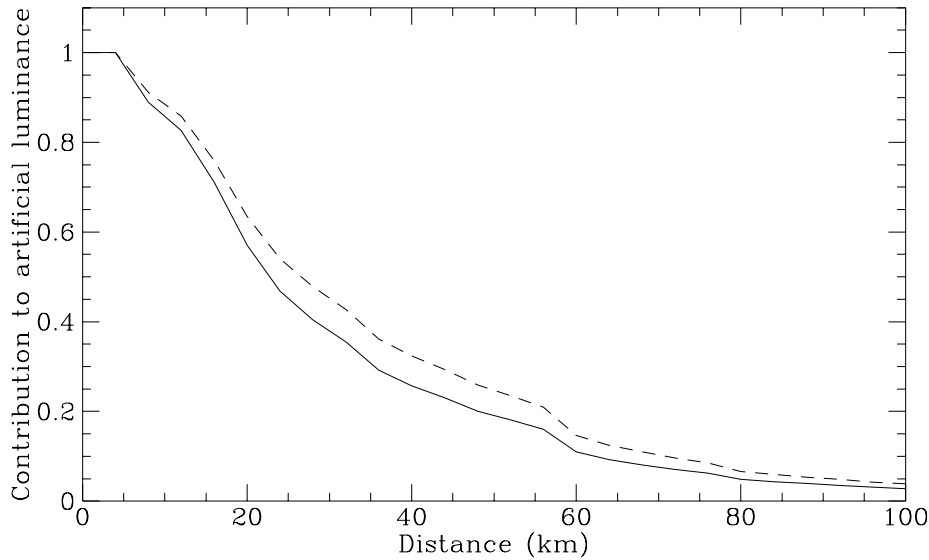} 
\caption{$b_d(d)$ curve for Mount Ekar Observatory.}
\label{fig3}
\end{figure}

\vspace{0.8cm} 

\centerline{\bf Tab. 2 - Vertical extinction predicted by the models}
\begin{table}[h]
\hspace{1.3cm} 
			\begin{tabular}{|l|r|r|r|r|}
			\hline
		Observatory  
		&  
		\multicolumn{2}{|c|}{K=1}& 	\multicolumn{2}{|c|}{K=2}\\
		&  
		\multicolumn{1}{|c}{$k_V$}& 	\multicolumn{1}{c|}{$k_B$}	&\multicolumn{1}{|c}{$k_V$}& 	\multicolumn{1}{c|}{$k_B$}\\				
\hline
Mount Ekar Obs. (Asiago)& 0.19 &0.36&0.24&0.43\\
Bologna Univ. Obs. (Loiano)& 0.24&0.43&0.34&0.56\\
Brera-Milan Ast. Obs. (Merate)& 0.28&0.50&0.42&0.66\\
Catania Obs. Stellar Sta. (Serra La Nave)& 0.16&0.32&0.20&0.38\\   
Collurania Ast. Obs. (Teramo)&0.28&0.48&0.40&0.64\\
Chaonis Obs. (Chions)& 0.33&0.56&0.50&0.78\\
``G. Ruggieri'' Obs. (Padova)& 0.33&0.56&0.50&0.78\\
S. Benedetto Po Ast. Obs. (S.Benedetto Po)&0.33&0.56&0.51&0.78\\
\hline
\end{tabular}
\end{table}

\begin{figure}
\epsfysize=8cm 
\hspace{3.5cm}\epsfbox{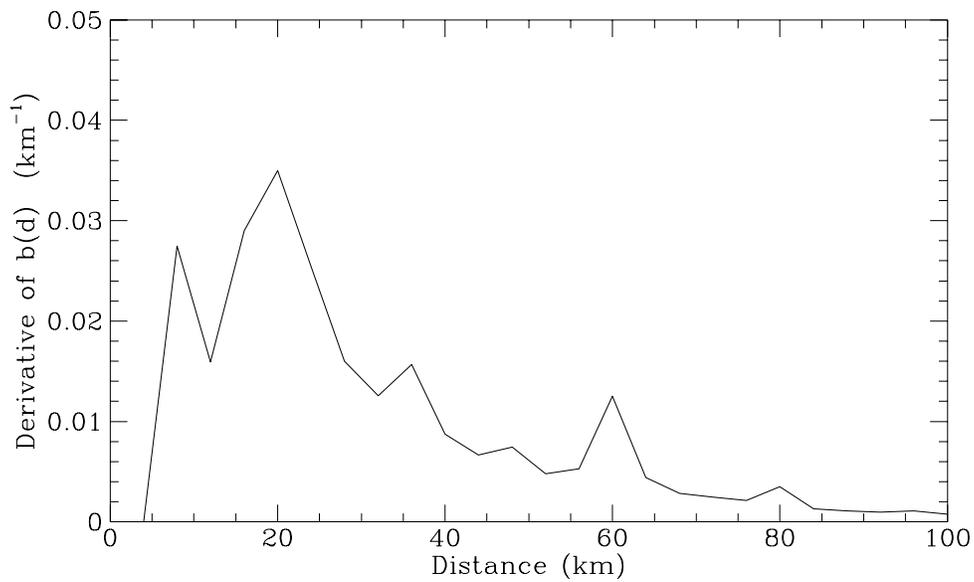} 
\caption{The derivative  $\frac{\partial b}{\partial d}$ of the $b_d(d)$ curve for Mount Ekar Observatory.}
\label{fig3b}
\end{figure}

Figure \ref{fig3} shows the $b_d(d)$ curve for the Mount Ekar Observatory.
The artificial sky luminance at the zenith of this
Observatory in clear nights is produced
mainly by the light dispersed from sources
situated in the Veneto plain. Approximately 50\% of the
artificial sky luminance at Ekar is produced within
30 km from the observatory and 75\% within 50 km.
As already shown in figure \ref{comp1} the shape of $b_d(d)$ is exactly that expected for a uniformly populated territory, except for a core in the inner 10 km produced by the lower population density in the mountain. The figure \ref{fig3b} shows the derivative  $\frac{\partial b}{\partial d}$. Main peaks correspond to the main sources like Asiago ( at $\sim5km$), Bassano, Thiene and Schio (from $10$ to $20km$), Vicenza (at $\sim35km$), Treviso (at $\sim56km$), Padova (at $\sim60km$). The continuum is produced by the other cities. I counted 1350 cities inside  
\clearpage

\begin{figure}
\epsfysize=8cm 
\hspace{3.5cm}\epsfbox{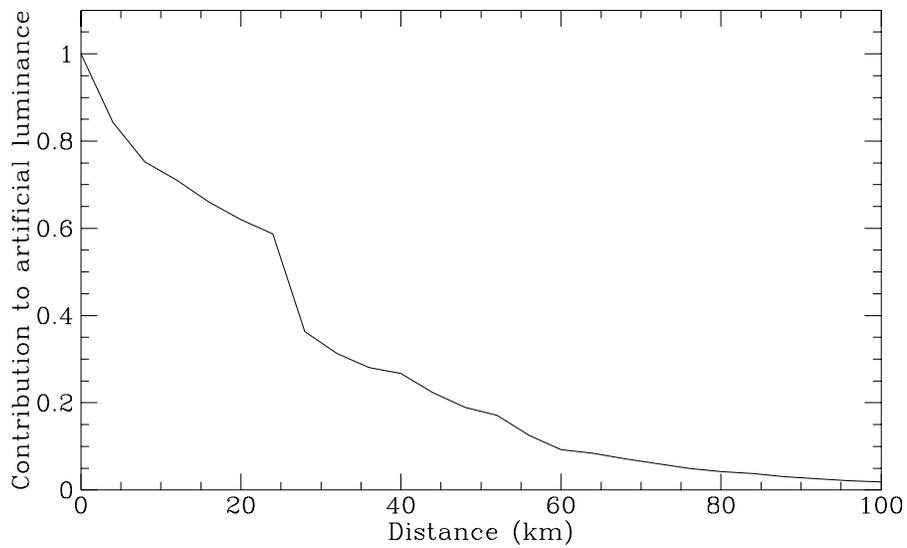} 
\caption{$b_d(d)$ curve for Bologna University Observatory at Loiano.}
\label{fig4}
\end{figure}

\clearpage

\noindent {100km from the site.}

Figure \ref{fig4} shows the $b_d(d)$ curve for Bologna University Observatory at Loiano. Readers can note the absence of the core due to the presence of the Loiano town near the Observatory and the strong bump at about 25 km produced by the big city of Bologna.
Approximately 50\% of the
artificial sky luminance at Loiano Observatory in clean nights 
is produced within
30 km from the observatory and 75\% within 50 km.

\begin{figure}
\epsfysize=8cm 
\hspace{3.5cm}\epsfbox{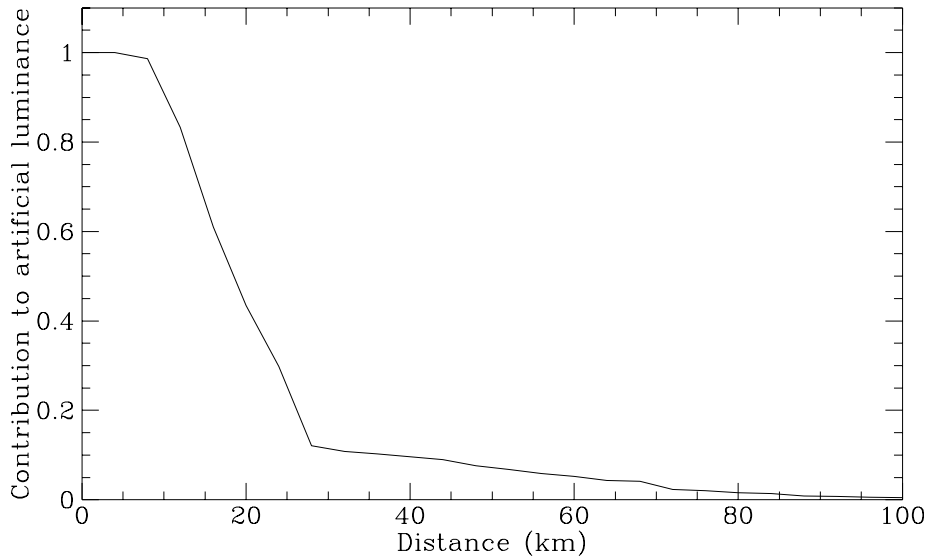} 
\caption{$b_d(d)$ curve for Catania Observatory Stellar Station at Serra La Nave.}
\label{fig6}
\end{figure}

\begin{figure}
\epsfysize=8cm 
\hspace{3.5cm}\epsfbox{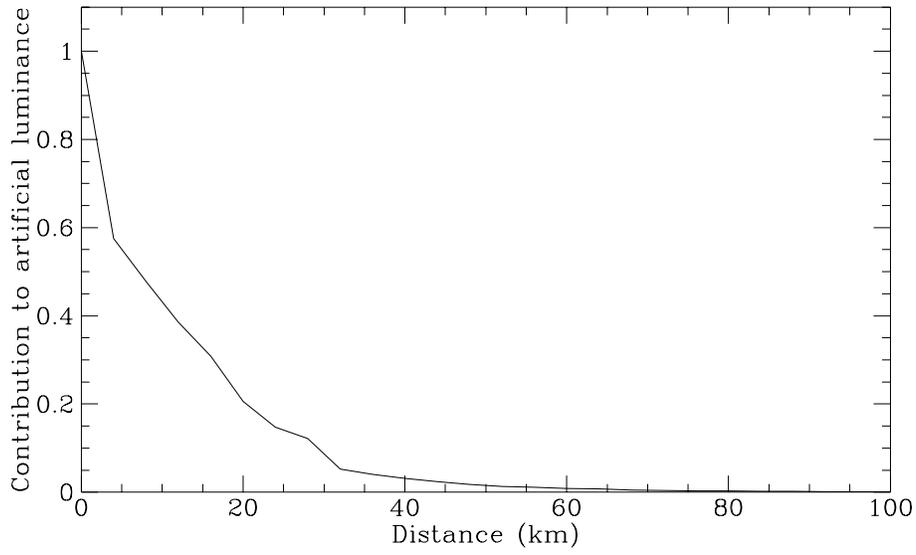} 
\caption{$b_d(d)$ curve for Brera-Milan Astronomical Observatory in Merate.}
\label{fig5}
\end{figure}

\begin{figure}
\epsfysize=8cm 
\hspace{3.5cm}\epsfbox{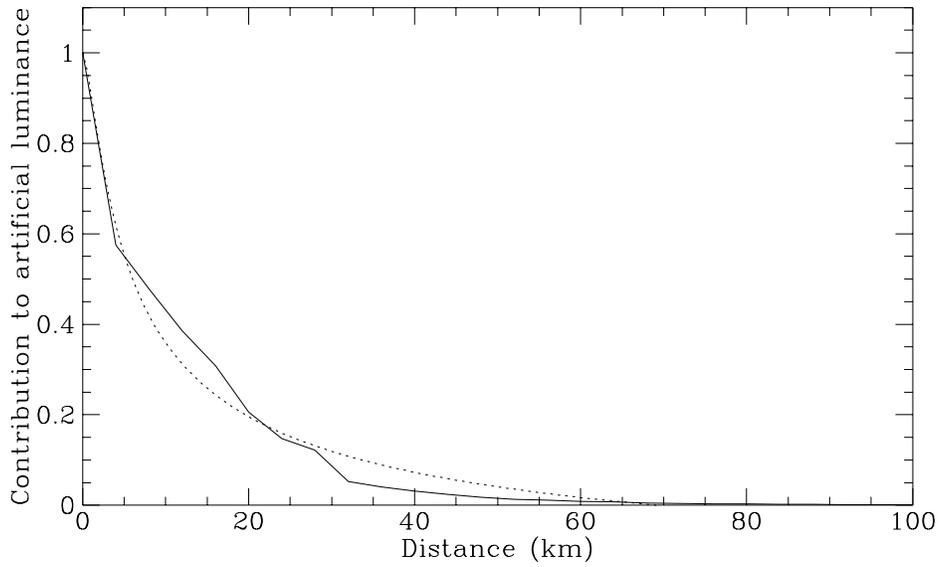} 
\caption{The  $b_d(d)$ curve for Brera-Milan Astronomical Observatory in Merate is well fitted by the $d^{-0.5}$ law.}
\label{fig5b}
\end{figure}

\begin{figure}
\epsfysize=8cm 
\hspace{3.5cm}\epsfbox{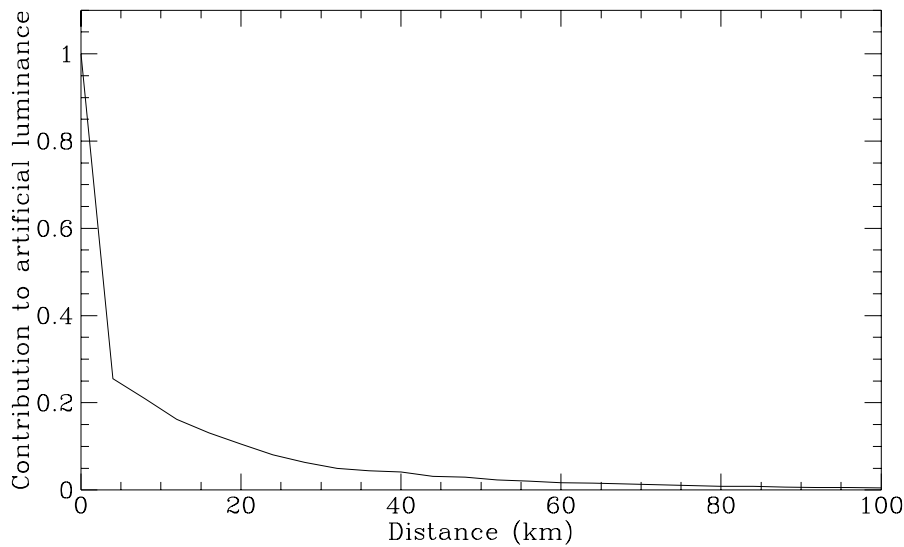} 
\caption{$b_d(d)$ curve for Collurania Observatory at Teramo.}
\label{fig10}
\end{figure}

It is interesting to compare the $b_d(d)$ curve of Catania Observatory Stellar Station at Serra La Nave in figure \ref{fig6}, Brera-Milano Astronomical Observatory at Merate in figure \ref{fig5} and Collurania Observatory at Teramo in figure \ref{fig10}.
Catania Observatory shows a large core due to the inhabitated Etna mountain and a rapid decrease between $\sim10 km$ to $\sim25 km$ produced by the fact that the city of Catania and its neighbour towns are the main contributors to the sky luminance at the site. Approximately 50\% of the
artificial sky luminance at Serra is produced within
30 km from the observatory and 75\% within 50 km. The contribution from higher distances is little due to the presence of the sea and to the absence of many other cities. Inside a radius of 120 km there are only 362 cities.
Brera-Milan Astronomical Observatory at Merate, on the contrary, do not show a core but a high peak because the observatory is almost surrounded from the city of Merate. The big number of cities in the land (2171 cities inside a circle of 120 km of radius around the site) produces the well prominent wing outside $\sim5 km$.
The diffuse population distribution produces a $b_d(d)$ curve well fitted by the extended $d^{-0.5}$ law (eq. \ref{bcontrib}) as shown in figure \ref{fig5b} ($\alpha=1.5$, $d_{c}=2.5$ km and $k$=0.12). The discontinuity at about 30 km in figure \ref{fig5} is produced from the contribution of the city of Milano.
A similar behaviour of $b_d(d)$ curve appear for the Collurania Observatory at Teramo where the contribution from the near city appears to overhang the contribution of the surrounding territory.
\begin{figure}
\epsfysize=8cm 
\hspace{3.5cm}\epsfbox{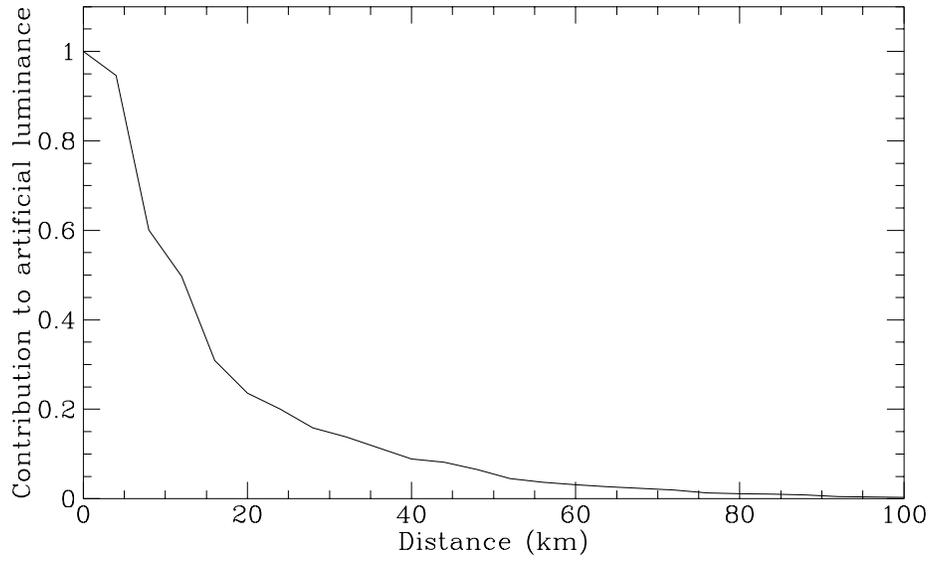} 
\caption{$b_d(d)$ curve for Chaonis Observatory.}
\label{fig9}
\end{figure}

\begin{figure}
\epsfysize=8cm 
\hspace{3.7cm}\epsfbox{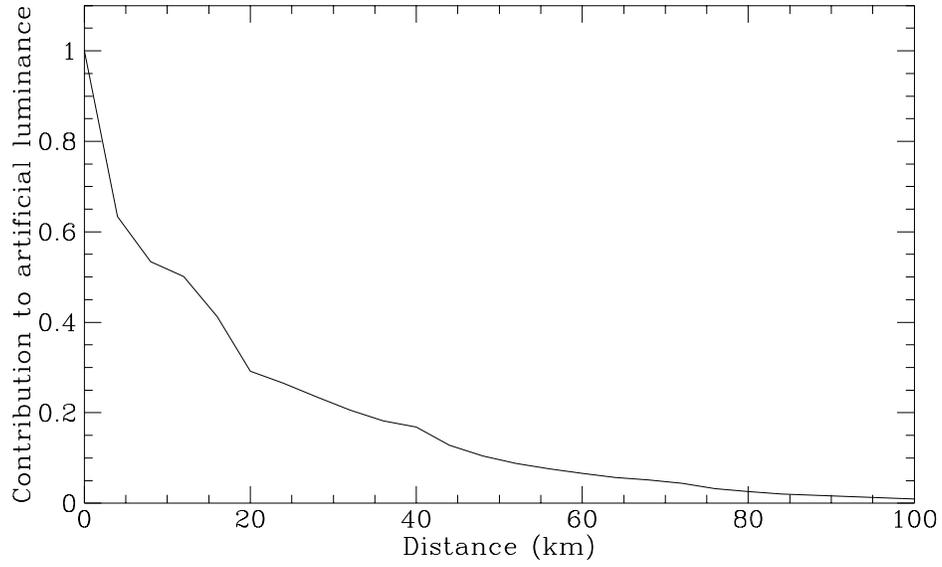} 
\caption{$b_d(d)$ curve for San Benedetto Po Observatory.}
\label{fig7}
\end{figure}

I also computed the $b_d(d)$ curve for three amateur observatories. Chaonis Observatory in figure \ref{fig9} shows a core of 5 km and a wing produced by the sources in the country, while the San Benedetto Po Observatory in figure \ref{fig7} shows a  wing but the core is replaced by a peak probably produced by the town in the nearby of the observatory. The bump at $\sim 16km$ is produced by the city of Mantova.
A little bump at $\sim40km$ is likely to be produced by the sum of the cities of Modena, Reggio Emilia and Verona.

\begin{figure}
\epsfysize=8cm 
\hspace{3.5cm}\epsfbox{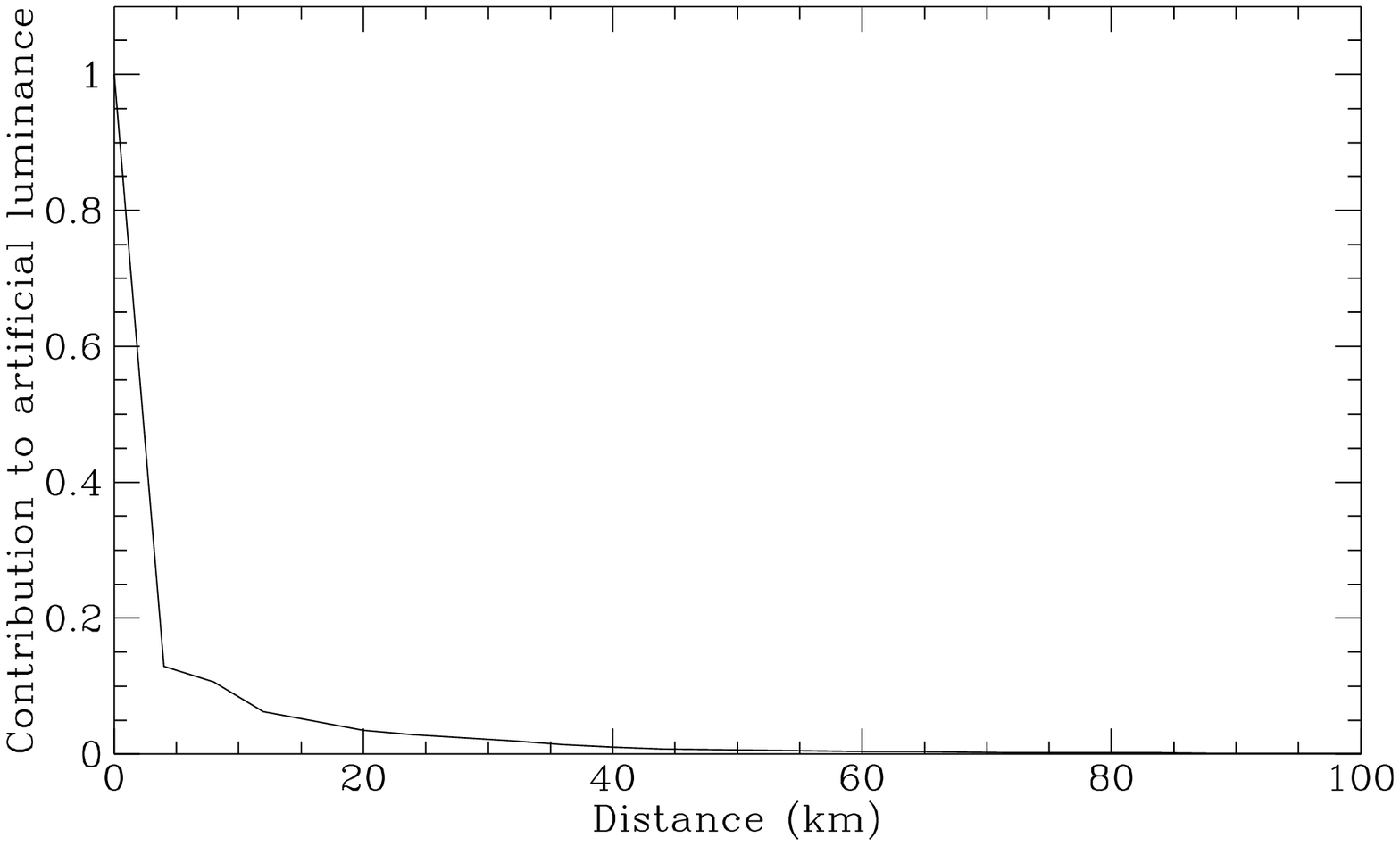} 
\caption{$b_d(d)$ curve inside the city of Padova.}
\label{fig8}
\end{figure}

\begin{figure}
\epsfysize=8cm 
\hspace{3.5cm}\epsfbox{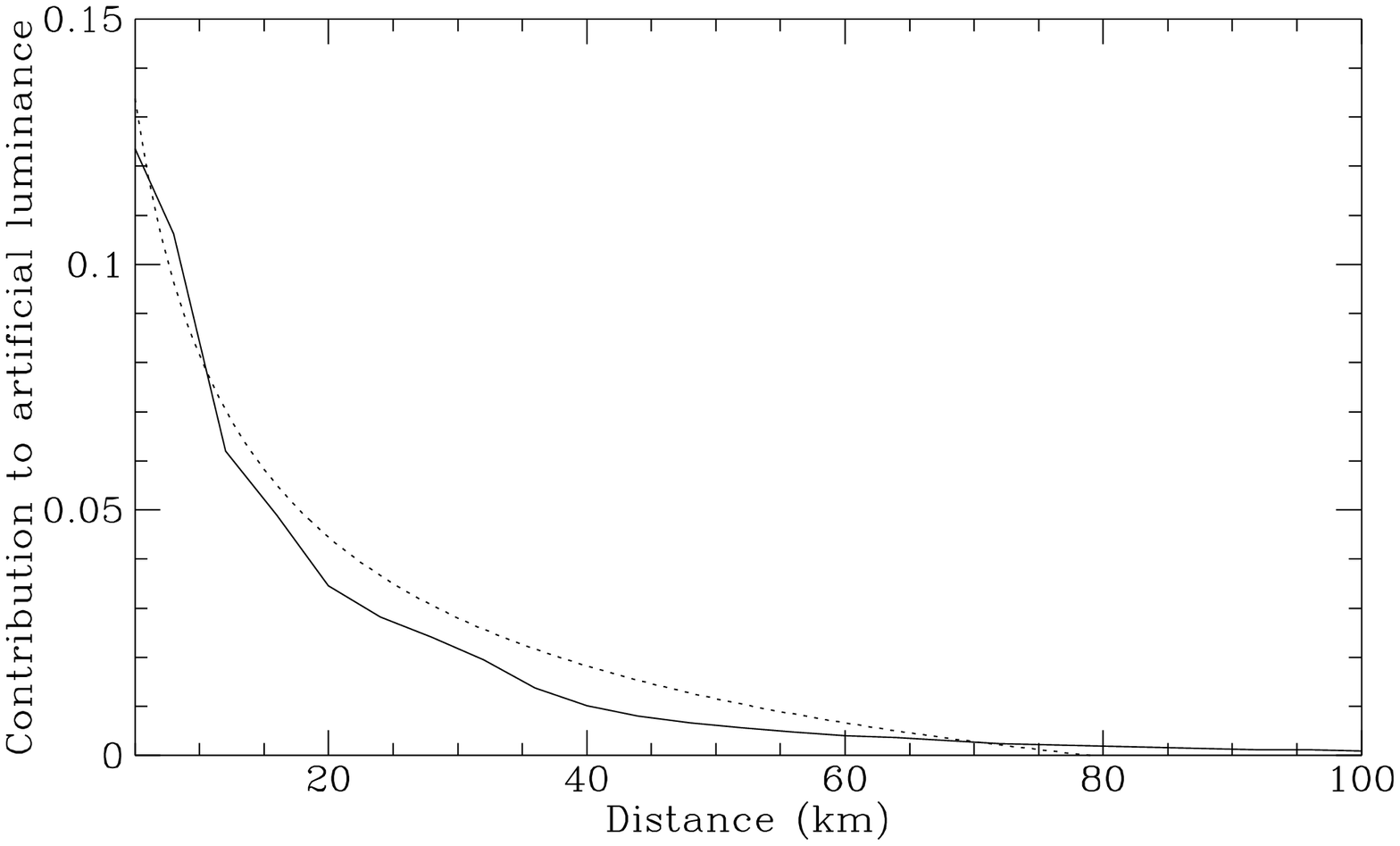} 
\caption{The outer wing of $b_d(d)$ curve for ``G. Ruggieri'' Observatory produced by the other cities of the Veneto plain is well fitted by the $d^{-0.5}$ law.}
\label{fig8b}
\end{figure}

The {\it G. Ruggieri} Observatory in figure \ref{fig8} is inside the city of Padova. The curve $b_d(d)$ shows the overhelming contribution coming inside the first 4 km. It is interesting to note that the wing produced by the other cities of the Veneto plain is quite well fitted by the $d^{-0.5}$ law ($\alpha=0$, $d_{c}=0.4$ km and $k$=0.045) as shown in figure \ref{fig8b}.

\section{Conclusions}
The knowledge of the   artificial
sky luminance $b_d(d)$ produced in a given point of the sky of a
site from the sources situated at a distance
greater than $d$ from the site
is important in order to understand the behaviour of light pollution in diffusely urbanized areas and it is useful in order to estimate which fraction
of the artificial luminance  would be regulated by norms or laws limiting the light wasted upward within protection areas of given radii. 

I  studied the behaviour of  $b_d(d)$ 
applying a model for the propagation of the light
pollution based on the modelling technique introduced by
Garstang (1986, 1987, 1988, 1989a, 1989b, 1989c, 1991a, 1991b, 1991c, 1992, 1993, 1999) which allows us to
calculate the  artificial luminance in a
given position on the sky of a site of given altitude on the sea level,
produced by a  source of given emission and geographic position.
I obtained $b_d(d)$ integrating
the contribution to the artificial luminance of
all the sources situated at a distance greater than $d$.

Main results are:
\begin{enumerate}
\item Artificial sky luminance in a site in a diffusely urbanised territory produced by sources located at large distances from the site is not negligible due at the addittive character of light pollution and its propagation at large distances.
\item The contribution $b_d(d)$ to the  artificial
sky luminance produced in given point of the sky of a
site by the sources situated at a distance
greater than a given distance $d$ from the site decrease as $ d^{-0.5}$ i.e. much slower than the contribution of a single source which goes as $d^{-2.5}$. 
\item With inclusion of a core, the expression (\ref{bcontrib}) well express the behaviour of $b_d(d)$ from 0 km to about 100 km.
\item The radii of protection zones around Observatories needs to be large in order that prescriptions limiting upward light be really effective.
\item Only when  the core radius is small, e.g. for Observatories located near a city, the sky luminance contribution from sources inside a small protection zone tends to be predominant.
\end{enumerate}

~~\\
In the Web Site  {\it Light Pollution in Italy}, actually at the address www.pd.astro.it/cinzano/ is presented a simple didactic Java Applet called LPCALC which allows to calculate $b_d(d)$ from the geographic position of a site. It works only for sites in Italy.

\acknowledgements
I am indebted to Roy Garstang of JILA-University of Colorado for his friendly kindness in reading and refereeing this paper, for his helpful suggestions and for interesting discussions.


\end{document}